\documentclass[multphys,vecphys]{svmult}

\usepackage{makeidx}     
\usepackage{graphicx}    
\usepackage{multicol}    

\makeindex             

\begin{document}

\title*{Supernova statistics}
\author{Enrico Cappellaro\inst{1} \and Roberto Barbon\inst{2} 
\and Massimo Turatto\inst{3}}
\institute{INAF - Osservatorio Astronomico di Capodimonte, via Moiariello 
16, 80181 Napoli (Italy) \texttt{cappellaro@na.astro.it} 
\and 
Dipartimento di Astronomia, Universit\`a di Padova, vicolo 
dell'Osservatorio 2, 35132 Padova (Italy) \texttt{barbon@pd.astro.it} 
\and 
INAF - Osservatorio Astronomico di Padova, vicolo 
dell'osservatorio 5, 35132 Padova (Italy) \texttt{turatto@pd.astro.it}}
%
%

\maketitle

\abstract

The statistics of SN discoveries is used to reveal selection
biases of past and current SN searches and to gain insight on the
progenitor scenarios for the different SN types. We also report
estimates of the SN rate per unit mass in galaxies of different types
and on the first attempts to study the evolution of the supernova rate
with redshift.

\section{Counting Supernovae}\label{why}

Event statistics is an invaluable tool to link the different supernova
types to their parent stellar populations and in turn to assess the
consistency of the possible progenitor scenarios.  Since the early
days of SN research, the simple observation that type Ia SNe are
found in all type of galaxies, including ellipticals where star
formation ceased long time ago, is used to deduce that their
progenitors must be low mass, long lived stars. Today the
standard scenario for type Ia calls for an accreting white dwarf in a
close binary system which explodes when it reaches the Chandrasekhar
mass. However, the real nature of the progenitor system has not yet
been identified and the different candidates, eg. double degenerate or
single degenerate, have still to pass the basic test of the occurrence
statistics \cite{Napi}.

Type II and Ib/c SNe are believed to be the outcomes of the core
collapse of stars with mass larger than 8-9 M$_\odot$. Different
sub-types have been related to progenitors with different initial
masses and/or metallicities \cite{Heger}. Since only for a few events
it has been possible to collect direct informations on the progenitors
\cite{Smartt}, one of the basic tool of investigation remains the
event statistics in systems with different stellar populations.

It is fair to say that the new interest for SNe in the last few years
has been driven not just by the wish to understand their physical
properties but mainly by their role as cosmological probes, in
particular the use of type Ia to measure the geometry of the Universe.
In addition, we believe that measurements of the SN rates as a function
of redshift is an attractive tool to recover the history of the star
formation rate with the cosmic age \cite{galtavi,sadat}.

\section{SN Searches}

\begin{figure}
\centering
\includegraphics[height=7cm]{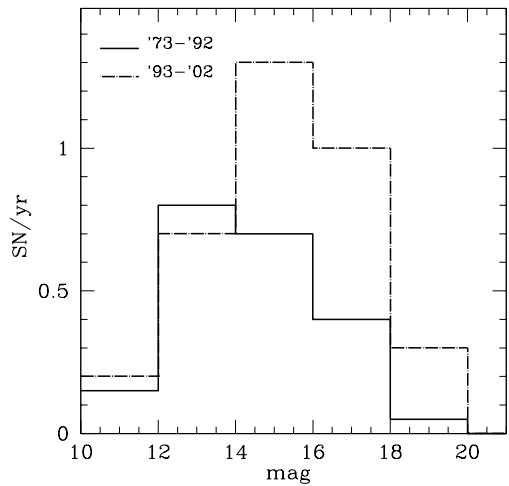}
\caption{Distribution of the apparent magnitude for the nearby SNe
(recession velocity $v_{hel}<1200\,{\rm km}\,{\rm s}^{-1}$) discovered
in two periods: the last decade (1993--2002) and the preceding twenty years
(1973--1992)}
\label{mag.fig}       
\end{figure}

At a rate of a few hundreds discovery per year, the number of known
SNe doubled in the last 5 years for a total count of over 
2500 events \footnote{See
http://www.pd.astro.it/supern/snean.txt for an up-to-date version of
{\em The Asiago SN Catalogue.}\cite{barbon}}. Yet, the discovery rate
of bright events, those $<15\,{\rm mag}$ at discovery, remains more or
less constant, at about 10 SN/year. This may appear to confirm the
common assumption that in modern time, say after 1970, all the SNe
which exploded in the local Universe were discovered \cite{tammann94}.
However, although most of the faint SNe discovered in the last decade
are distant events, there is also a significant contribution for
nearby SNe. This can be seen in Fig.~\ref{mag.fig} where we compare
the apparent magnitude distribution of the SNe discovered in
the local Universe ($v_{hel}<1200\,{\rm km}\,{\rm s}^{-1}$) in two
different periods, the last decade and the twenty years before.  It
turns out that the discovery rate in nearby galaxies is today almost a
factor 2 larger than in the past and that the average apparent
magnitude at discovery is about 1 mag fainter.

It is important to stress that among the faint, nearby events there
are not only those SNe that, due to seasonal observational
limitations, were discovered long after maximum, but also highly
extinguished events (eg. SN~2002cv \cite{valentini}) and intrinsically
faint SNe (eg. SN~1997D \cite{tur97d}).

The presence of such bias explains why the absolute value of the SN
rate cannot be derived from the general list of SNe but instead through a
detailed analysis of the data of an individual SN search,
for which actual limits and biases can be carefully accounted for.

\begin{figure}
\centering
\includegraphics[height=10cm]{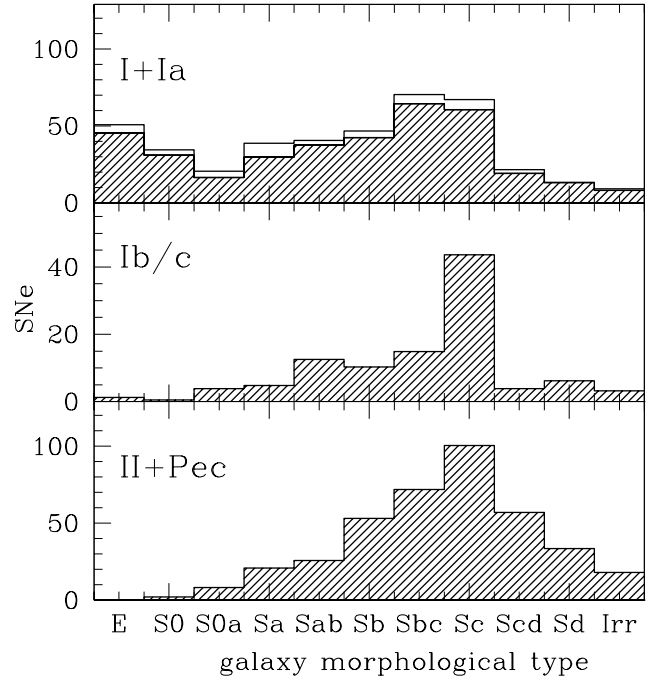}
\caption{SN counts for galaxies of different 
morphological type for the three main SN types, normalised to the
fraction of galaxy of the given type as derived from the RC3 galaxy
catalog \cite{rc3}.  Only SNe at redshift $z<0.05$ have been included. In the
upper panel the shaded area are type Ia SNe only}
\label{gtype.fig}       
\end{figure}

The main problem, when considering the data of an individual SN search,
is that the statistics is not very large, especially when considering
less frequent SN types. Therefore the analysis of the relative SN
rates from the general list can still be of interest.  In
Fig.~\ref{gtype.fig} we report, in a separate panel for each of the
three main SN type (Ia, II and Ib/c), the SN counts as a function of
galaxy type. It results that the discovery rate of type Ia SN appears
more or less independent on galaxy type, while the core collapse SN
rate rapidly increases from early to late type spirals. The latter
gives a very close match of the current estimates of the star
formation rate in galaxies of different type (cf. Fig.3 in
\cite{kennicutt}) and directly reflects the fact that SN~II derive
from massive, short lived progenitors. An intriguing feature which is
seen in Fig.~\ref{gtype.fig} is the spike for type Ib/c SNe in Sc
galaxies which, because of a similar peak in the star formation rate,
suggests that, in the average, their progenitors are more massive than
those of normal type II SNe.  We should note that this effect seems to
be washed-out when using, instead of the RC3 galaxy classification
system as in the Asiago SN catalogue, the DDO system \cite{vdb}.

Because of the peak of the SFR, the ratio of the events in Sc galaxies
compared to the total counts for a given SN type can be used to
rank the SN types according to their progenitor masses. This ratio
results $12\pm1$\% for type Ia, $19\pm2$\% for type II and
$34\pm6$\% for type Ib/c which gives the same indication of Fig.~2 in a
different form and can be used as reference to derive some hints
for less frequent SN sub-types. For instance, we found that the same
ratio for type IIn is $24\pm6$\% which suggests that these SNe have in
average the same progenitor mass, or maybe even somewhat higher, than
normal type II. This has become interesting after the discovery of SN
2002ic \cite{hamuy}, a SN~Ia showing evidence of interaction with
a dense H envelope and, at late time, developing a spectrum very similar to that
of some type IIn like SN 1997cy.  That this is the most common
channel for type IIn is not consistent with the fact that their
progenitors, in the average, seems to be massive (cf. \cite{vdb}).

Another example is that of faint type II SNe \cite{tur97d}. These low
energy explosions delivering one order of magnitude less Ni that
normal type II have been related to the formation of a black hole
rather than to a neutron star. This is attributed to their
progenitors being more massive than those of normal SNII
\cite{zampieri,pastorello} and it is consistent with the fact that out of
10 events, 5 occurred in Sc galaxies.

\begin{figure}
\centering
\resizebox{10cm}{!}{\rotatebox[]{-90}{\includegraphics{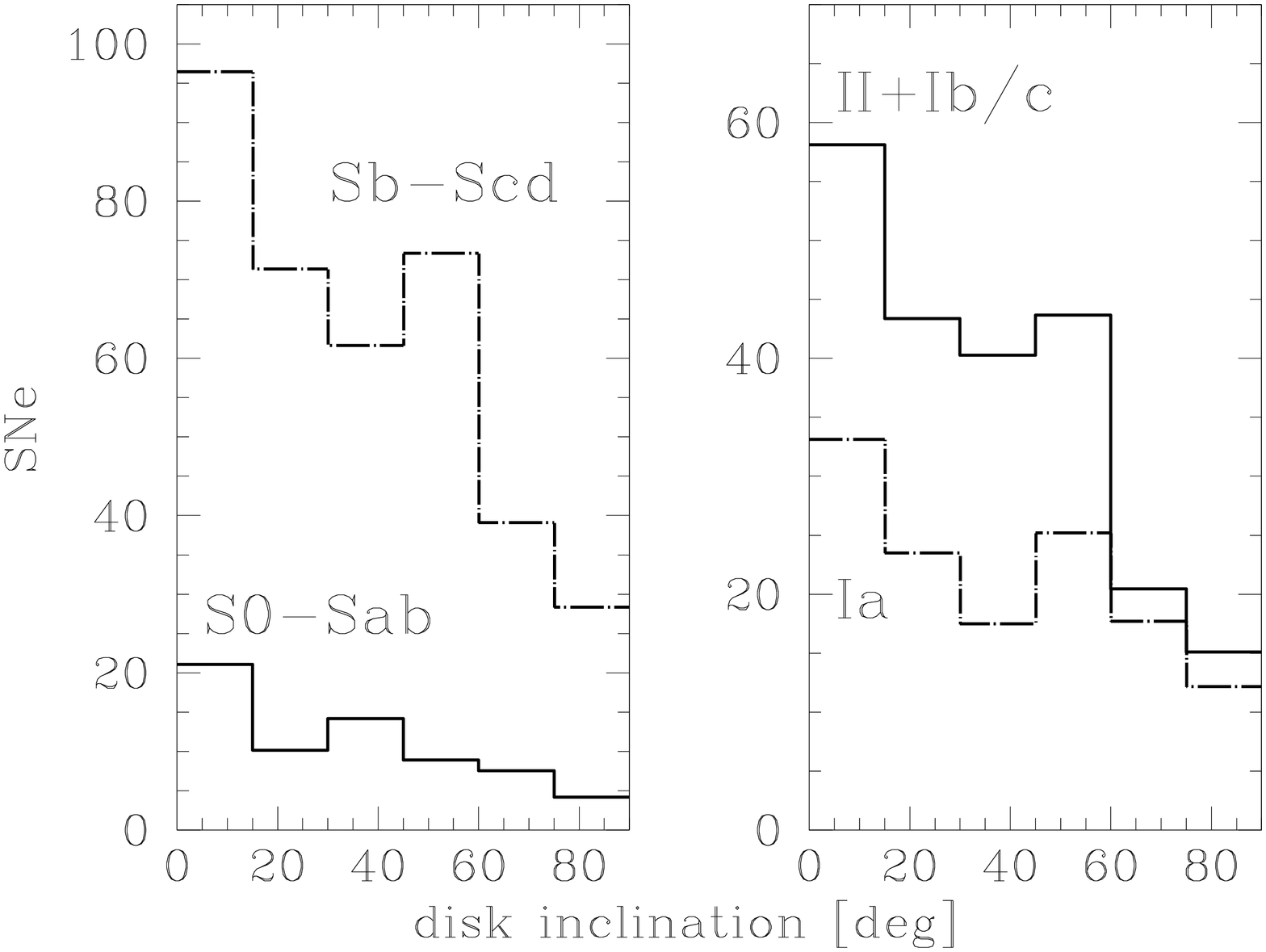}}}
\caption{SN counts in spiral galaxies of different inclinations (0 deg
is for face-on galaxies) normalized to the fraction of galaxies in each
bin of inclination as given in the RC3. In the left panel we show
separately early and late type spirals, whereas in the right panel we
distinguish core collapse (II+Ib/c) from thermonuclear (Ia)
Supernovae. Only the SNe discovered after 1998 have been included.}
\label{fig:inc}       
\end{figure}

\section{SN Rates}

As we mentioned before, the absolute value of the SN rates requires that the
detection efficiency and selection biases of individual SN searches are
accurately estimated. At the same time, it is important to maintain a
sufficient statistics for SN events which, especially in the past, was
not achievable using one search alone.  An obvious solution is to pool
the data from a few systematic SN searches as in \cite{stat99}.  
Somewhat surprisingly, these most significant results still
relay on photographic and/or visual SN searches whereas
the legacy of the modern systematic CCD SN searches have still to be
exploited.

One of the main uncertainties on the current estimate of the SN rates
is the bias against SN detection in spiral galaxies which are not seen
face-on. Likely, this is  due to the fact that the larger optical depth
trough the dust, and hence the higher extinction, makes the average SN
in inclined spirals fainter that in face-on ones. It is often claimed
that the effect was only important for photographic surveys whereas
modern CCD searches, due to the better spectral sensitivity in the red,
are less affected. Although this is certainly true, the inclination
effect does remain important. This is shown in Fig.\ref{fig:inc},
where the SN discovery counts in spiral galaxies of different
inclination are compared for different grouping of galaxy and SN
types. Only SNe discovered in the last 5 years, hence mainly in
systematic CCD searches, have been included.   It
turns out that taking as reference the discovery rate in face-on
spirals ($i<30^\circ$), even in the present day SN searches we are
missing half of the type Ia and 2/3 of the core collapse SNe occuring
in edge-on spirals ($i>60^\circ$).
We stress that although
it may occur that in some SN searches face-on galaxies are monitored
more frequently, it is not expected that this depends on galaxy or SN type. 

After Tammann \cite{tammann}, the rate of SNe is usually normalized to
the galaxy blue luminosity (${\rm SNu} = {\rm SN}\,10^{-10}{\rm
L}_{\odot}^{\rm B}\,10^{-2}{\rm yr}$). This is convenient because $a)$
the luminosity can easily be measured for large galaxy samples and $b)$
the SN rate has been found to be proportional to the galaxy
luminosity.  However, while interpreting the results, it has to be taken into
account that there are different contributors to a galaxy blue luminosity
depending on its stellar population mixture.  Indeed it is well know
that the mass to light ratio ($M/L$) changes by one order of magnitude
along the Hubble galaxy sequence. We can convert SN rate per unit
luminosity to SN rate per unit mass assuming an average $M/L$ for each
galaxy type \cite{faber}. The results are reported in
Tab.~\ref{tab:rate}.

It is known been that for SN~Ia the rate per unit luminosity remains
almost constant moving from ellipticals to spirals. On the other side
as it can be seen from Tab.~1, because M/L is lower, the rate of SNIa per unit mass in
late spirals is almost 3 times higher than in ellipticals.  This implies
that a fraction of SN Ia in spirals must be related to a relatively
young stellar population.

Because of the short time scale of evolution of massive stars, the
core collapse (II+Ib/c) SN rate can be translated in the present time
star formation rate if we know the initial mass function and the mass
range of core collapse progenitors. Assuming a Salpeter mass function
(with index 1.35) and choosing 8 and 40 M$\odot$ for the lower and upper limits of
core collapse progenitor masses, from the measured core-collapse SN rates we derive
than in a
$10^{11}\,$M$_\odot$ Sbc-Sd galaxy of the local Universe the star
formation rate is 1.8 M$_\odot\,$yr$^{-1}$, in S0a-Sb is 0.7
M$_\odot\,$yr$^{-1}$, whereas for E-S0 we derive an upper limit of $0.01$
M$_\odot\,$yr$^{-1}$. These numbers strongly depend on the
lower mass limit for core collapse progenitors which is not well
known. For instance if we take for the latter 10 M$_\odot$ the
estimate of the star formation rates increases by $\sim 40$\%.

\begin{table}
\centering
\caption{SN rate per unit mass [$10^{-11}{\rm M}_\odot\,10^{-2}{\rm yr}\, \times \,({\rm H}/75)^2$].}
\label{tab:rate}       
%
%
\begin{tabular}{lcccc}
\hline\noalign{\smallskip}
galaxy& \multicolumn{4}{c}{SN rate }\\
\cline{2-5}
type~~~~~~~~~~~~~    &  ~~~~~Ia~~~~~~~~~     &  ~~~~~~~Ib/c~~~~~~~    & ~~~~~~~~II~~~~~~~~ & ~~~~~~~~~~~All~~~~~~~~~~\\
\noalign{\smallskip}\hline\noalign{\smallskip}
E-S0    &  $0.16\pm0.03$ & \multicolumn{2}{c}{$<0.01$}  & $0.16\pm0.03$\\
S0a-Sb  &  $0.29\pm0.07$ & $0.16\pm0.07$ & $0.69\pm0.17$& $1.14\pm0.20$ \\
Sbc-Sd &   $0.46\pm0.10$ & $0.30\pm0.11$ & $1.89\pm0.34$& $2.65\pm0.37$ \\
All    &   $0.27\pm0.03$ & $0.11\pm0.03$ & $0.53\pm0.07$& $0.91\pm0.08$ \\
\noalign{\smallskip}\hline
\end{tabular}
\end{table}

\section{Evolution of the SN rate with redshift.}

Estimates of the supernova rate, in particular that of core
collapse SNe at different redshifts can be used to recover the history
of star formation with cosmic age. Conversely, if the latter is known,
it is possible to constrain the SN progenitor scenario which is
especially important for type Ia which has
not yet been firmly established.  Despite of these prospectives, very few
observational estimates have been published to date and all rely on
the major efforts devoted to the search of type Ia to be used as
cosmological distance indicators. This means that $a)$ only type Ia
rate have been measured and $b)$ the search strategy introduces severe
biases.  In particular, to allow for accurate photometry, the
candidates found in the galaxy inner regions and/or in bright galaxies
are usually rejected.

A further concern is that little is known about the properties of the
galaxy sample.  Among other things, this makes impossible to verify
the presence of a spiral inclination effect such as in
the local Universe. We stress that although high redshift SN searches
are usually performed in the red (R or I band), the host galaxy
extinction occurs in the SN rest frame wavelength 
(ie. B or V for $z\sim0.5$). Therefore if in galaxies at $z\sim0.5$
the dust content and properties are the same as in the local
Universe, we expect similar biases.

The few available estimates seem to indicate an evolution of the
type Ia SN rate with redshift. At $z=0.55$ the measured value is
$0.33^{+0.06}_{-0.05} (H/75)^2$ SNu \cite{pain} to be compared with
the local estimate (not corrected for host galaxy inclination) of
$0.14\pm0.04 (H/75)^2$ \cite{stat99}. Taken as a face value, this
indicates a very rapid increase of the SN Ia rate.  However, due to
the large uncertainty, this is still consistent with the current
scenario of galaxy and SN Ia progenitor system evolution
\cite{kobayashi}.

In the last few years we started a long term project which is
especially designed to measure SN rates and to overcome some of
the previous limitations.  In particular we try to:

\begin{enumerate}
\item reduce as much as possible the candidate selection biases. Indeed
we do count candidates in the galaxy nuclear region (but
we try to reduce the contamination from variable AGN using the long term
variability history of the source).
\item count both type Ia and core collapse SNe. Core collapse SN rate
can be used to constraints the star formation rate at the given
redshift. Besides, relative SN rates have smaller systematic errors
than absolute values.
\item use photometric redshift to characterize the galaxy sample
\end{enumerate}

The first results of this program are now becoming available
\cite{galtavi} and indicate that the rate of core collapse SNe grows
at a faster pace than SN Ia. In particular we found
evidence that the star formation rate a $z=0.30$ is a factor 3 higher
than in the local Universe. Although this is a preliminary result it
confirms similar finding based on the measurements of the
H$\alpha$ emission from galaxies at the same redshift \cite{fujita}.
The coming in operation of new wide field telescopes (eg. VST+OmegaCAM
\cite{vst,kuijken} and LBT+LBC \cite{roberto} will give the chance to
strongly improve the statistics and to probe different redshifts.

%
%

%
%



\printindex
\end{document}